\documentclass[final,5p,times,twocolumn]{elsarticle}

\usepackage{amsmath}
\usepackage{amssymb}
\usepackage{graphicx}
\usepackage{bm}
\usepackage{braket}
\usepackage{enumitem}
\usepackage{tensor}
\usepackage{combelow}
\usepackage{slashed}
\usepackage{braket}
\usepackage{tensor}
\usepackage{color}

\def\psibar{{\overline{\psi}}}
\def\tr{{\ensuremath{\text{tr}}}}

\journal{Physics Letters B}

\begin{document}

\begin{frontmatter}

\title{Renormalised fermion vacuum expectation values on anti-de Sitter space-time}

\author[arft]{Victor E. Ambru\cb{s}}
\ead{Victor.Ambrus@gmail.com}
\author[uos]{Elizabeth Winstanley}
\ead{E.Winstanley@sheffield.ac.uk}
\address[arft]{Center for Fundamental and Advanced Technical Research,
Romanian Academy, Bd.~Mihai Viteazul 24, Timi\cb{s}oara 300223, Romania}
\address[uos]{Consortium for Fundamental Physics, School of Mathematics and Statistics, University of Sheffield,\\
Hicks Building, Hounsfield Road, Sheffield S3 7RH, United Kingdom}

\date{\today}

\begin{abstract}
The Schwinger--de Witt and Hadamard methods are used to obtain renormalised
vacuum expectation values for the fermion condensate, charge current and stress-energy tensor of a quantum fermion field of
arbitrary mass on four-dimensional anti-de Sitter space-time. The quantum field is in the global anti-de Sitter vacuum state. The results are compared with
those obtained using the Pauli-Villars and zeta-function
regularisation methods, respectively.
\end{abstract}


\end{frontmatter}

\section{Introduction}

Quantum field theory on anti-de Sitter space-time (adS) has been of particular interest since the formulation of the
adS/CFT (conformal field theory) correspondence (see \cite{art:aharony00} for a review).
The maximal symmetry of adS simplifies many aspects of the study of quantum fields on this background.  For instance,
closed-form expressions for the Feynman propagator for a quantum field in the global adS vacuum state can be derived
for both bosonic \cite{art:allen86} and fermionic fields \cite{art:lutken86,art:muck00}.
In quantum field theory on curved space-time, an object of fundamental importance is the renormalised expectation value
of the stress-energy tensor, $\braket{T_{\mu\nu}}$, since this governs the backreaction of the quantum field on the
space-time geometry. Recently, Hadamard renormalisation \cite{art:dec08} has been used to give closed-form expressions
for the renormalised vacuum expectation value of the stress-energy tensor for a quantum scalar field with arbitrary
coupling to $n$-dimensional adS \cite{art:kent14}.

In this letter,
we study a quantum fermion field $\psi $ on four-dimensional adS space-time and consider the global vacuum state.
Vacuum expectation values  (v.e.v.s) of the stress-energy tensor for such a fermion field have been computed previously
\cite{art:camporesi92} using Pauli-Villars and zeta-function regularisation. Our purpose in this paper is to compare those
results with v.e.v.s calculated using two alternative approaches to renormalisation, namely
the Schwinger--de Witt method \cite{art:christensen78} and the Hadamard method \cite{art:najmi84,art:dappiaggi09}.
For both approaches, we find the v.e.v.s of the fermion condensate (FC) $\braket{\psibar \psi}$,
the charge current (CC) $\braket{J^\mu}$ and the stress-energy tensor (SET) $\braket{T_{\mu\nu}}$
when the fermion field is in the global adS vacuum state.

\section{Dirac equation on adS}

We consider the vacuum state of Dirac fermions of arbitrary mass $m$ on a
four-dimensional adS background space-time of inverse radius of curvature
$\omega$ (so that the Ricci scalar curvature is $R = -12\omega^2$).
The space-time metric takes the form
\begin{equation}
ds^{2} = \frac {1}{\cos ^{2} \left( \omega r \right) } \left[ -dt^{2} + dr^{2}
+ \frac {\sin ^{2} \left( \omega r \right) }{\omega ^{2}} \left( d\theta ^{2} + \sin ^{2} \theta \, d\varphi ^{2} \right) \right] ,
\label{eq:metric}
\end{equation}
where we use the metric signature $\left( - , + , + , +\right) $ and units in which $G=c=\hbar=1$.

The Dirac equation for a fermion field of mass $m$
on an arbitrary space-time can be written as:
\begin{equation}
 (i \gamma^\mu D_\mu - m) \psi(x) = 0,
 \label{eq:dirac}
\end{equation}
where the point-dependent gamma matrices $\gamma^\mu \equiv \gamma^\mu(x)$ satisfy
generalised canonical anti-commutation relations $ \{\gamma^\mu, \gamma^\nu\} = -2 g^{\mu\nu}$
with $g^{\mu\nu}$ the inverse of the metric $g_{\mu\nu}$.
The spinor covariant derivatives $D_\mu$ are defined to ensure the covariance of the Dirac
equation with respect to general coordinate transformations.
On the adS metric \eqref{eq:metric}, using a suitable choice of tetrad basis vectors, mode solutions of the Dirac equation can be found
\cite{art:cot07}.
In this letter, we study only
v.e.v.s with respect to the global adS vacuum state,
for which the Feynman propagator can be found using a geometric approach.
We therefore do not consider mode solutions of the Dirac equation.

\section{Feynman propagator for the global adS vacuum}

For a general quantum state, the Feynman propagator for the fermion field, $S_F(x,x')$, can be determined as a solution of the inhomogeneous Dirac equation:
\begin{equation}
\left( i \slashed{D} - m \right) S_F(x,x') = \frac{1}{\sqrt{-g}} \delta^4(x - x'),
\label{eq:sf_eq}
\end{equation}
where the Feynman slash notation is used to denote contractions with the gamma matrices $\gamma ^{\mu }$
(e.g.~$\slashed{D}=\gamma ^{\mu }D_{\mu }$).

The maximal symmetry of the global adS vacuum state allows $S_F(x,x')$ for this state to be written as \cite{art:muck00}:
\begin{equation}
i S_F(x,x') = (\alpha_F + \beta_F \slashed{n}) \Lambda(x,x'),
\label{eq:sf_muck}
\end{equation}
where $\alpha_F$ and $\beta_F$ are scalar functions of the geodetic interval $s$ and
the tangent at $x$ to the geodesic connecting $x$ and $x'$ is denoted by  $n_\mu(x,x') = \nabla_\mu s(x,x')$.
We use the convention that $s(x,x')$ is real if the geodesic connecting $x$ and $x'$ is
time-like or null, implying a negative norm for $n_\mu$, that is $ g_{\mu\nu} n^\mu n^\nu = -1$.
The bispinor of parallel transport $\Lambda (x,x')$ along the geodesic connecting $x$ and $x'$ satisfies the
parallel transport equation for spinors at both ends of the geodesic \cite{art:muck00}:
\begin{equation}
 n^\mu D_\mu \Lambda(x,x') = 0, \qquad n^{\mu'} \Lambda(x,x') \overleftarrow{\overline{D}}_{\mu'} = 0,
 \label{eq:ptL}
\end{equation}
together with the initial condition $ \left.\Lambda(x,x')\right\rfloor_{x' = x} = 1$.
In the equations above,
unprimed and primed indices denote quantities evaluated with respect to $x$ and $x'$, respectively.
An overbar denotes the Dirac adjoint.

On adS, the geodesic tangent $n_\mu$ and bispinor of parallel transport $\Lambda$ satisfy the following
equations \cite{art:allen86,art:muck00}:
\begin{subequations}\label{eq:nablas}
\begin{align}
 \nabla_\nu n_\mu =& -\omega  \left( g_{\mu\nu} + n_\mu n_\nu \right) \cot \left( \omega s \right) ,
 \label{eq:nablan}
 \\
 \nabla_{\nu'} n_\mu =& -\frac{\omega}{\sin \left( \omega s \right)} \left( g_{\mu\nu'} - n_\mu n_{\nu'} \right),
 \label{eq:nablanp}\\
 D_\mu \Lambda(x,x') =& \omega \Sigma_{\mu\nu} n^\nu \Lambda(x,x')
 \tan \left( \frac{\omega s}{2} \right) ,
 \label{eq:nablaL}
\end{align}
\end{subequations}
where
$\Sigma _{\mu \nu} = \frac {1}{4} \left[ \gamma _{\mu }, \gamma _{\nu } \right] $
are the anti-Hermitian generators of Lorentz transformations and
$g_{\mu\nu'} \equiv g_{\mu\nu'}(x,x')$ is the bivector of parallel transport, which satisfies
the parallel transport equations:
\begin{equation}
 n^\lambda \nabla_\lambda g_{\mu\nu'} = 0, \qquad
 n^{\lambda'}\nabla_{\lambda'} g_{\mu\nu'} = 0.
\end{equation}

Substituting the ansatz \eqref{eq:sf_muck} into \eqref{eq:sf_eq} and using the above properties
of $n_\mu$ and $\Lambda (x,x')$, the inhomogeneous Dirac equation can be reduced to two decoupled equations:
\begin{subequations}
\label{eq:sf_dec}
\begin{align}
 i \alpha_F' - \frac{3i\omega}{2} \alpha_{F} \tan \left( \frac{\omega s}{2} \right) - m \beta_F =& 0,
 \label{eq:ads_beta_from_alpha}
 \\
 i \beta_F' + \frac{3i\omega}{2} \beta _{F} \cot \left( \frac{\omega s}{2} \right) - m \alpha_F =&
 \frac{i}{\sqrt{-g}} \delta(x,x').
\end{align}
\end{subequations}
These two equations can be combined to form a single second order differential equation for $\alpha_F$:
\begin{multline}\label{eq:alpha_eq}
 \frac{\partial^2 \alpha_F}{\partial(\omega s)^2} + 3 \cot \left( \omega s \right) \frac{\partial \alpha_F}{\partial(\omega s)}
 + \left[k^2 -
 \frac{3}{4 \cos^2 \left(\frac{\omega s}{2}\right)} - \frac{9}{4}\right] \alpha_F\\
 = -\frac{ik}{\omega} \frac{\delta(x - x')}{\sqrt{-g}},
\end{multline}
where
\begin{equation}
 k = \frac{m}{\omega}.
 \label{eq:kdef}
\end{equation}
Changing variable to $z = \cos^2\left( \frac{\omega s}{2} \right) $ and writing $\alpha_F = z^{\frac{1}{2}} \widetilde{\alpha}_F$
puts \eqref{eq:alpha_eq} in the form of the hypergeometric equation (in agreement with \cite{art:muck00}):
\begin{multline}\label{eq:alpha_eqz}
 \left[z(1-z) \frac{d^2}{dz^2} + (3-5z) \frac{d}{dz} + (k-2)(k+2)\right] \widetilde{\alpha}_F\\
 = -\frac{ik}{\omega \sqrt{z}} \frac{\delta(x-x')}{\sqrt{-g}}.
\end{multline}

For the calculation of v.e.v.s, it is useful to express $S_F(x,x')$ in terms of
a quantity that goes to $0$ in the coincidence limit (i.e.~as $s\rightarrow 0$).
Writing \eqref{eq:alpha_eqz} in terms of $q = \sin^2\left( \frac{\omega s}{2} \right) $
gives:
\begin{multline}
 \left[q(1-q) \frac{d^2}{dq^2} + (2-5q) \frac{d}{dq} - (2-k)(2+k)\right] \widetilde{\alpha}_F\\
 = -\frac{ik}{\omega \sqrt{1-q}} \frac{\delta(x-x')}{\sqrt{-g}}.
 \label{eq:alpha_eqq}
\end{multline}
This has the form of a hypergeometric differential equation with parameters $a = 2 -k$, $b = 2+k$ and $c = 2$,
and its general solution can be written in terms of two arbitrary constants $\lambda $, $\lambda '$ as \cite{book:nist}:
\begin{multline}
\label{eq:alphaq}
 \alpha_F = \lambda  \left\{- \frac{1}{(k^2-1)q}  + {}_2F_1(2-k,2+k;2;q) [\lambda' + \ln(-q)]  \right. \\
 \left.+ \sum_{n=0}^\infty \frac{(2+k)_n (2-k)_n}{(2)_n n!} q^n \Psi_n\right\}
 \cos \left( \frac{\omega s}{2}  \right) ,
\end{multline}
where
$(z)_n = z(z+1)\dots(z+n - 1)$
is the Pochhammer symbol \cite{book:nist,book:asteg}
and
\begin{equation}
 \Psi_n = \psi(2+k+n) + \psi(2-k+n) - \psi(2+n) - \psi(1+n)
 \label{eq:Psin}
\end{equation}
is defined in terms of the polygamma function $\psi(z) = d \left[ \ln\Gamma(z) \right] / dz$.

The constant $\lambda$ can be found by matching the small distance behaviour of $\alpha_F$ with
that of the Minkowski quantity $\alpha_{\text{Mink}}$
\cite{book:birrell}
\begin{equation}
 \alpha_{\text{Mink}} = \frac{im^2}{8\pi s} H^{(2)}_1 (m s)
 \mathop {\sim }_{s\rightarrow 0} - \frac{m}{4\pi^2 s^2},
 \label{eq:alpha_s0}
\end{equation}
where $H^{(2)}_{1}$ is a Hankel function of the second kind.
Thus, $\lambda$ is given by:
\begin{equation}
\label{eq:muck_aq_lambdadiv}
 \lambda = \frac{k \omega^3}{16\pi^2} \left( k^2 - 1 \right).
\end{equation}

To fix the remaining constant $\lambda '$, we consider large spatial separations of the points $x$ and $x'$.  With our conventions, the geodetic
interval $s$ is purely imaginary for space-like separated points, so we set $s=i{\tilde {s}}$, where ${\tilde {s}}$ is real.
Using properties of the hypergeometric functions \cite{book:asteg}, the function $\alpha _{F}$ can be rewritten in the form \cite{thesis:victor}
\begin{multline}
\alpha _{F} = \frac {\omega ^{3}\Gamma \left( 2 + k \right)}{16 \pi ^{\frac {3}{2}} 4^{k} \Gamma \left( \frac {1}{2} + k \right) }
\cosh \left( \frac {\omega {\tilde {s}}}{2} \right) \left[
\sinh ^{2} \left( \frac {\omega {\tilde {s}}}{2} \right) \right] ^{-2-k} \\
\times {}_{2}F_{1}\left( 1+k, 2+k; 1+2k ; - {\text{cosech}} ^{2} \left( \frac {\omega {\tilde {s}}}{2} \right) \right) \\
+ \frac {\omega ^{3}k\left( k^{2}-1 \right) }{16\pi ^{2}} \left[ \lambda ' + \pi \cot \left( \pi k \right) \right]
\cosh \left( \frac {\omega {\tilde {s}}}{2} \right) \\
\times {}_{2}F_{1} \left( 2+k, 2-k ; 2 ; -\sinh ^{2} \left( \frac {\omega {\tilde {s}}}{2} \right) \right) .
\label{eq:a_z}
\end{multline}
For very large spatial separations (as one of the points $x$, $x'$ moves close to the adS boundary), the first hypergeometric function in the
expression for $\alpha _{F}$ \eqref{eq:a_z} is regular and multiplied by a very small factor.
However, the second hypergeometric function is divergent.
Therefore, in order for $\alpha _{F}$ to remain finite for large spatial separations, it must be the case that
\begin{equation}
\label{eq:lambdap}
 \lambda' = -\pi \cot \left( \pi k \right).
\end{equation}
Thus, the function $\alpha_F$ takes the form \eqref{eq:alphaq} with $\lambda '$ given by \eqref{eq:lambdap}.
The function $\beta _{F}$ is found from $\alpha _{F}$ \eqref{eq:alphaq} using \eqref{eq:ads_beta_from_alpha} and is given by
\begin{align}
 \beta_F =& -\frac{i \omega^3 k^2 (k^2 - 1)}{16\pi^2}\left\{
 -\frac{1}{k^2(k^2 -1 )q^2} (1 + k^2 q)\nonumber\right.\\
 &+ \frac{1}{2} \left[- \pi \cot \left(\pi k \right) + \ln (-q)\right]
 {}_2F_1\left(2+k,2-k;3;q\right) \nonumber\\
 &\left. + \frac{1}{2} \sum_{n=0}^\infty \frac{(2+k)_n (2-k)_n}{(3)_n n!} q^n
 \left(\Psi_n - \frac{1}{2+n}\right)\right\}  \sin\left( \frac{\omega s}{2}\right) .
 \label{eq:b_q}
\end{align}

\section{Vacuum expectation values}

Vacuum expectation values (v.e.v.s) of the fermion condensate (FC) $\braket{\psibar \psi}$, charge current (CC) $\braket{J^\mu}$
and stress-energy tensor (SET) $\braket{T_{\mu\nu}}$ are calculated from the Feynman propagator derived in the previous section
using the following results \cite{art:groves02}:
\begin{subequations}
\label{eq:vevs}
\begin{align}
 \braket{\psibar \psi} =& -\lim_{x'\rightarrow x} \tr \left[ i S_F(x,x')
 \Lambda(x',x)\right],\label{eq:ppsi_sf}\\
 \braket{J^\mu} =& -\lim_{x'\rightarrow x} \tr\left[\gamma^\mu i S_F(x, x')
 \Lambda(x',x)\right], \label{eq:jmu_sf}\\
 \braket{T^{\text{can}}_{\mu\nu}} =& \frac{i}{2} \lim_{x'\rightarrow x} \tr\left\{
 \left[\gamma_{(\nu} D_{\mu)} iS_F(x, x') -
 iS_F(x,x')\overleftarrow{\overline{D}}_{\lambda'} \gamma_{\kappa'} g\indices{^{\lambda'}_{(\mu}}
 g\indices{^{\kappa'}_{\nu)}}
 \right]\right.\nonumber\\
 &\left. \times\Lambda(x',x)\right\}.
 \label{eq:Tmunu_sf}
\end{align}
\end{subequations}
The superscript on $T^{\text{can}}_{\mu\nu}$ indicates that the canonical definition for the stress-energy
tensor operator is used.
If the ansatz \eqref{eq:sf_muck} is made for the form of $S_F(x,x')$,
equations \eqref{eq:nablan} and \eqref{eq:nablaL}
can be used to write the above v.e.v.s using the functions $\alpha_F$ and $\beta_F$:
\begin{subequations}
\label{eq:ads_vevs_ab}
\begin{align}
 \braket{\psibar \psi} =& 4 \lim_{x'\rightarrow x} \alpha_F(s),
 \label{eq:ads_ppsi_ab}\\
 \braket{J^\mu} =& 4 \lim_{x'\rightarrow x} n^\mu \beta_F(s),
 \label{eq:ads_jmu_ab}\\
 \braket{T_{\mu\nu}^{\text {can}}} =& 4i\lim_{x'\rightarrow x} \left\{
 -n_\mu n_\nu \left[\frac{\partial}{\partial s} - \frac{\omega}{2} \cot \left(\frac{\omega s}{2}\right) \right] \beta_F \right. \nonumber \\
&\left. +
 g_{\mu\nu} \frac{\omega}{2} \beta_F \cot\left(\frac{\omega s}{2} \right) \right\}.
 \label{eq:ads_set_ab}
\end{align}
\end{subequations}
The tangents to the geodesic $n_\mu$ depend on the direction along which the points are split. For consistency,
their coefficients should cancel
in the coincidence limit, since the final expressions for the
v.e.v.s above must be independent of the point-splitting employed. Thus, the v.e.v.~of the CC $\braket{J^\mu}$
will vanish identically.
Furthermore, the adS symmetries imply that the v.e.v.~of the FC $\braket{\psibar \psi}$ must be a constant scalar,
while the v.e.v.~of the SET is a constant multiplying the metric tensor $g_{\mu\nu}$, such that
\begin{equation}
 \braket{T_{\mu\nu}^{\text{can}}}_{\text{ren}} = \frac{1}{4} g_{\mu\nu} \braket{T^{\text{can}}}_{\text{ren}},
\end{equation}
where $T$ is the trace of $T_{\mu\nu}$.

The above expressions \eqref{eq:ads_vevs_ab} are all infinite due to the divergence of $\alpha _{F}$ and $\beta _{F}$
in the coincidence limit $s\rightarrow 0$.
This divergence can be seen clearly in the small $s$ behaviour of $\alpha _{F}$ \eqref{eq:alpha_s0} and $\beta _{F}$:
\begin{subequations}\label{eq:ab}
\begin{align}
 \alpha_F =& -\frac{k\omega}{4\pi^2s^2} -
 \frac{\omega^3}{16\pi^2} \left(1 + \frac{5k}{6} - k^2 + k^3\right)\nonumber\\
 &+ \frac{\omega^3 k(k^2 - 1)}{8\pi^2} \left[\psi(k) + \gamma +
 \frac{1}{2} \ln\left(-\frac{\omega^2 s^2}{4}\right)  \right] + O(s^2),\\
 \beta_F =& \frac{i}{2\pi^2s^3} + \frac{i\omega^2(1+2k^2)}{16\pi^2s}\nonumber\\
 & + \frac{i\omega^4 s}{64 \pi^2} \left(\frac{17}{60} + k +
 \frac{5k^2}{6} - k^3 + \frac{3k^4}{2}\right)\nonumber\\
 &- \frac{i\omega^4 s}{32\pi^2} k^2(k^2 - 1) \left[\psi(k) + \gamma +
 \frac{1}{2} \ln\left(-\frac{\omega^2 s^2}{4}\right)  \right] + O(s^3),
\end{align}
\end{subequations}
where $\gamma $ is Euler's constant.

\section{Regularisation of the Feynman propagator}

The renormalisation of the v.e.v.s in \eqref{eq:ads_vevs_ab} can be performed by
regularising $\alpha_F$ and $\beta_F$, as follows:
\begin{equation}
 \alpha_F^{\text{reg}} = \alpha_F - \alpha_{\text{div}}, \qquad
 \beta_F^{\text{reg}} = \beta_F - \beta_{\text{div}},
 \label{eq:ab_div}
\end{equation}
such that $\alpha_F^{\text{reg}}$ and $\beta_F^{\text{reg}}$
stay finite as $s\rightarrow 0$.
The singularity structure of the fermion propagator can be studied by considering the
auxiliary propagator $\mathcal{G}_F$, defined as \cite{art:najmi84}:
\begin{equation}
 S_F(x,x') = \left( i \slashed{D} + m \right) \mathcal{G}_F(x,x').
 \label{eq:g_def}
\end{equation}
From the inhomogeneous Dirac equation \eqref{eq:sf_eq}, the auxiliary propagator $\mathcal{G}_F(x,x')$ satisfies
the following equation \cite{art:najmi84}:
\begin{equation}
 \left(\Box - \tfrac{1}{4} R - m^2\right) \mathcal{G}_F(x,x') = \frac{1}{\sqrt{-g}} \delta^4(x,x'),
 \label{eq:gf_eq}
\end{equation}
where $R$ is the Ricci scalar curvature.
It should be emphasised that the
auxiliary propagator $\mathcal {G}_{F}$, like the Feynman propagator $S_{F}$,
is a bispinor. Therefore, the
box operator above is understood to be written in terms of spinor covariant derivatives.
If the Feynman propagator $S_{F}$ has the form \eqref{eq:sf_muck},
then \eqref{eq:nablaL} can be used to show that:
\begin{equation}
 i\mathcal{G}_F(x,x') = \frac{\alpha_F(s)}{k \omega} \Lambda(x,x').
 \label{eq:gf_alpha}
\end{equation}
Thus, the divergent part of the propagator can be written as:
\begin{equation}
 i\mathcal{G}_{\text{div}}(x,x') = \frac{\alpha_{\text{div}}}{k \omega} \Lambda(x,x').
 \label{eq:gdiv_alpha}
\end{equation}
In the following sections, we use the Schwinger--de Witt and Hadamard regularisation methods
to find ${\mathcal {G}}_{\text{div}}$ and hence the renormalised v.e.v.s.

\section{Schwinger--de Witt renormalisation}
\label{sec:sdw}

Using the Schwinger--de Witt expansion, Christensen \cite{art:christensen78}
finds the following expression for $\mathcal{G}_{\text{div}}^{\text{SdW}}$:
\begin{multline}
 i\mathcal{G}_{\text{div}}^{\text{SdW}} = \frac{\sqrt{\Delta}}{8\pi^2} \Bigg\{
 a_0 \left[\frac{1}{\sigma} + m^2\left(1 + \frac{m^2 \sigma}{4}\right) L - \frac{m^2}{2} - \frac{5}{16} m^2 \sigma\right]\\
 - a_1\left[\left(1 + \frac{m^2 \sigma}{2}\right) L - \frac{m^2 \sigma}{2}\right] +
 a_2 \sigma \left[\left(\frac{1}{2} + \frac{m^2\sigma}{8}\right) L - \frac{1}{4}\right]\Bigg\},
 \label{eq:sdw_gf}
\end{multline}
where $\sigma = -\frac{s^2}{2}$, the Van Vleck-Morette determinant is denoted $\Delta $ and
\begin{equation}
 L = \gamma + \frac{1}{2}\ln \left(\frac{\nu^2_{\text{SdW}} \sigma}{2}\right)
\end{equation}
is written in terms of an arbitrary renormalisation mass scale
$\nu_{\text{SdW}}$.
The coefficients $a_n$ are bispinors regular as $s \rightarrow 0$ and satisfy the
following differential equation \cite{art:christensen78}:
\begin{equation}
 \sigma^\rho a_{n + 1; \rho} + (n + 1) a_{n + 1} = \frac{1}{\sqrt{\Delta}} \left(\Box - \frac{R}{4}\right)
 \left(\sqrt{\Delta} a_n\right),\label{eq:sdw_eq_anp1}
\end{equation}
with $a_0$ given for any space-time by \cite{art:christensen78}:
\begin{equation}
 a_0(x,x') = \Lambda(x,x').
\end{equation}

Specialising to adS, the functions $a_n$ can be written as
\begin{equation}
 a_n(x,x') = \omega^{2n} A_n(s) \Lambda(x,x'),
\end{equation}
where $A_{n}(s)$ are scalar functions.
Using the explicit expression
\begin{equation}
 \Delta(x,x') = \left(\frac{\omega s}{\sin\omega s}\right)^3
 \label{eq:vanvleck}
\end{equation}
for the Van Vleck-Morette determinant on adS \cite{art:kent14},
together with \eqref{eq:nablas}, the differential equation \eqref{eq:sdw_eq_anp1} reduces to:
\begin{multline}
 -\frac{1}{(\omega s)^n} \frac{\partial}{\partial(\omega s)} \left[(\omega s)^{n + 1} A_{n+1}\right]\\
 = \frac{\partial^2 A_n}{\partial(\omega s)^2} + \frac{3}{\omega s}\frac{\partial A_n}{\partial (\omega s)} -
 \frac{3}{4}\left[\frac{1}{\sin^2\omega s} - \frac{1}{(\omega s)^2} + \frac{1}{\cos^2\frac{\omega s}{2}}\right]A_n.
\end{multline}
The above equation can be solved as a power series in $\omega s$, yielding:
\begin{equation}
A_{0}=1, \quad
A_{1} = 1 + \frac{19}{240} (\omega s)^2 + O(s^4), \quad
A_{2}= \frac{11}{60} + O(s^2).
\end{equation}
Combining \eqref{eq:gdiv_alpha} and \eqref{eq:sdw_gf} gives
$\alpha_{\text{div}}^{\text{SdW}}$ as a power series:
\begin{subequations}\label{eq:sdw_ab_div}
\begin{multline}
 \alpha_{\text{div}}^{\text{SdW}} = -\frac{k\omega}{4\pi^2 s^2} - \frac{\omega^3 k (1+k^2)}{16\pi^2}
 - \frac{k\omega^5 s^2}{64\pi^2} \left(\frac{9}{20} + 3k^2 - \frac{5k^4}{4}\right)\\
 + \frac{\omega^3 k (k^2 - 1)}{8\pi^2} \left[1 + \frac{3 - k^2}{8} (\omega s)^2\right] L + O(s^4).
 \label{eq:sdw_a_div}
\end{multline}
We construct $\beta_{\text{div}}^{\text{SdW}}$ by using $\alpha_{\text{div}}^{\text{SdW}}$ in \eqref{eq:ads_beta_from_alpha}:
\begin{multline}
 \beta_{\text{div}}^{\text{SdW}} = \frac{i}{2 \pi^2 s^3} + \frac{i\omega^2(1 + 2k^2)}{16\pi^2s} +
 \frac{i\omega^4 s}{64\pi^2}\left(\frac{1}{10} + k^2 + \frac{3k^4}{2} \right)\\
 -\frac{i\omega^4 s}{32\pi^2} k^2 (k^2 - 1) L + O(s^3).
 \label{eq:sdw_b_div}
\end{multline}
\end{subequations}

Subtracting \eqref{eq:sdw_ab_div} from \eqref{eq:ab} and substituting the result
in \eqref{eq:vevs} gives the following renormalised v.e.v.s:
\begin{subequations}
\begin{align}
 \braket{\psibar\psi}_{\text{ren}}^{\text{SdW}} =& -\frac{\omega^3}{4\pi^2}\left(1 - \frac{k}{6} - k^2\right)
\nonumber\\&
 + \frac{\omega^3 k (k^2 - 1)}{2\pi^2} \left[\psi(k) + \ln \left(\frac{\omega}{\nu_{\text{SdW}}}\right)\right],
 \label{eq:sdw_ppsi}\\
 \braket{J^\mu}_{\text{ren}}^{\text{SdW}} =& 0,\label{eq:sdw_jnu}\\
 \braket{T^{\text{can}}}_{\text{ren}}^{\text{SdW}} =&
 -\frac{\omega^4}{4\pi^2}\left(\frac{11}{60} + k - \frac{k^2}{6} - k^3\right)\nonumber\\
 &+ \frac{\omega^4 k^2(k^2 - 1)}{2\pi^2} \left[\psi(k) + \ln\left( \frac{\omega}{\nu_{\text{SdW}}}\right)\right].
 \label{eq:sdw_tren}
\end{align}
\end{subequations}
When $k=0$, we find the expected trace anomaly for massless fermion fields:
\begin{equation}
 \braket{T}_{k=0} = -\frac{11\omega^4}{240\pi^2}.
 \label{eq:traceanomaly}
 \end{equation}

The result \eqref{eq:sdw_tren} can be compared with the trace $\braket{T}_{\text{ren}}^{\text{P-V}}$ of the
SET obtained using Pauli-Villars regularisation \cite{art:camporesi92}:
\begin{multline}\label{eq:pv_tren}
 \braket{T}_{\text{ren}}^{\text{P-V}} = -\frac{\omega^4}{4\pi^2}\left(\frac{11}{60} + k - \frac{k^2}{6} - k^3\right)\\
 + \frac{\omega^4 k^2(k^2 -1)}{2\pi^2} \left[\psi(k) + \ln\left(\frac{\omega}{\nu_{\text{P-V}}}\right)\right],
\end{multline}
where $\nu_{\text{P-V}}$ is an arbitrary renormalisation mass scale. The agreement between
\eqref{eq:sdw_tren} and \eqref{eq:pv_tren} is excellent, provided that the two renormalisation
mass scales $\nu _{\text {SdW}}$ and $\nu _{\text {P-V}}$ are equal,
$ \nu_{\text{P-V}} = \nu_{\text{SdW}}$.

\section{Hadamard renormalisation}\label{sec:had}

\subsection{Hadamard form}

Hadamard renormalisation is a mathematically rigorous approach to regularisation of v.e.v.s
(see \cite{art:dappiaggi09,thesis:Kohler,art:kratzert00,art:hollands01,art:sahlmann01} for mathematical details
for the fermion case).
The divergent part of the auxiliary propagator \eqref{eq:g_def} is known as the Hadamard form $\mathcal{G}^{\text{Had}}_{\text {div}}$.
This is purely geometric, depending on the space-time background but not the quantum state under consideration.
The remainder, $\mathcal{G}_{\text{reg}}(x,x')={\mathcal {G}}_{F}(x,x') - \mathcal {G}^{\text{Had}}_{\text {div}}(x,x')$,
is regular as $x' \rightarrow x$ and depends on the quantum
state of the fermion field.

The Hadamard form $\mathcal{G}^{\text{Had}}_{\text {div}}(x,x')$ of the auxiliary propagator
can be written as \cite{art:najmi84,art:dappiaggi09}:
\begin{equation}
 i\mathcal{G}^{\text {Had}}_{\text {div}}(x,x') = \frac{1}{8\pi^2} \left[\frac{U(x,x')}{\sigma} +
 V(x,x')\,\ln(\nu_{\text{Had}}^2\sigma)\right] ,
 \label{eq:uv_def}
\end{equation}
where $\nu_{\text{Had}}$ is an arbitrary renormalisation mass scale.
The bispinors $U(x,x')$ and $V(x,x')$ are regular in the coincidence
limit $x'\rightarrow x$ and are determined by the following equations, which follow
from substituting \eqref{eq:uv_def} into \eqref{eq:gf_eq}:
\begin{subequations}
\label{eq:UVeqns}
\begin{align}
 \sigma^\lambda U_{;\lambda} + \tfrac{1}{2}(\Box \sigma - 4) U =& 0
 \label{eq:U},\\
 \sigma^\lambda V_{;\lambda} + \tfrac{1}{2}(\Box \sigma - 2) V
 + \tfrac{1}{2}(\Box - \tfrac{1}{4} R - m^2) U =& O(\sigma),
 \label{eq:UV}\\
 (\Box - \tfrac{1}{4} R - m^2)V =& 0 .
 \label{eq:V}
\end{align}
\end{subequations}
We emphasise that all covariant derivatives in \eqref{eq:UVeqns} are spinor covariant derivatives.

On a four-dimensional space-time, the solution of \eqref{eq:U} can be found analytically \cite{art:najmi84}:
\begin{equation}
 U(x,x') = \sqrt{\Delta(x,x')} \Lambda(x,x').
\end{equation}
Since $V(x,x')$ satisfies
the homogeneous version \eqref{eq:V}
of \eqref{eq:gf_eq} which governs the auxiliary propagator,
the symmetries of adS allow $V(x,x')$ to
be put in a form similar to  \eqref{eq:gf_alpha}:
\begin{equation}
 V(x,x') = \frac{\alpha_V(s)}{k \omega} \Lambda(x,x').
\end{equation}
Here $\alpha_V(s)$ is the solution of the homogeneous version of \eqref{eq:alpha_eqq} which is regular at the origin:
\begin{equation}
 \alpha_V(s) = k\omega\,\mathcal{C}\,  {}_2F_1\left(2-k,2+k;2;q\right) \cos\left( \frac{\omega s}{2} \right).
 \label{eq:alpha_V_sol}
\end{equation}
The integration constant $\mathcal{C}$ is fixed by imposing \eqref{eq:UV}.
It can be seen that the first term in \eqref{eq:UV} is of order $\sigma$. The second term evaluates to:
\begin{equation}
 \frac {1}{2} \left[ \Box \sigma - 2 \right] V = \frac {1}{2}\left[ -1 + 3 \omega s \cot  \left( \omega s \right) \right] V
 = {\mathcal {C}} \Lambda + O(\sigma ),
 \label{eq:boxsigma}
\end{equation}
while the third term can be shown to equal:
\begin{multline}
 \frac{1}{2}\left(\Box - \frac{1}{4} R - m^2\right) U\\
 = \frac{3\omega^2}{8} \left(-\frac{1}{(\omega s)^2}
 + \frac{1}{\sin^2 \left( \omega s \right) } + \frac{1}{\cos^2\left( \frac{\omega s}{2}\right) } - \frac{4k^2}{3}\right) U
 \\
 = - \frac {\omega ^{2}}{2} \left(k^2 - 1\right) \Lambda +O(\sigma ).
 \label{eq:box_U}
\end{multline}
Therefore, the integration constant in $\alpha_V$ is given by
\begin{equation}
\mathcal{C} = \frac{\omega^2}{2} (k^2 - 1) .
\end{equation}

From the derivation of $U$ and $V$ above, we find that
\begin{multline}
\label{eq:alphaH}
 \alpha^{\text {Had}}_{\text {div}} = \frac{k \omega}{8\pi^2}\Bigg\{\frac{\sqrt{\Delta}}{\sigma} + \frac{\omega^2}{2} (k^2 - 1)
 _2F_1\left(2-k,2+k;2;q\right) \\
\times \cos\left( \frac{\omega s}{2} \right)
 \ln(\nu_{\text{Had}}^2 \sigma)\Bigg\}.
\end{multline}
Equation~\eqref{eq:ads_beta_from_alpha} can be used to find $\beta^{\text {Had}}_{\text {div}}$:
\begin{multline}
 \beta^{\text {Had}}_{\text {div}} = \frac{i\omega^3}{8\pi^2}\Bigg\{\frac{\sqrt{\Delta}}{(\omega s)^3} +
 \frac{3\sqrt{\Delta}}{(\omega s)^2 \sin \left( \omega s\right) }\\
  - \frac{1}{4}k^2(k^2 - 1) {}_2F_1(2-k,2+k;3;q)
  \sin \left(\frac{\omega s}{2}\right) \ln(\nu_{\text{Had}}^2 \sigma)   \\
  + \frac{k^2 - 1}{\omega s}
 {}_2F_1(2-k,2+k;2;q) \cos\left(\frac{\omega s}{2} \right)
  \Bigg\}.
\end{multline}

\subsection{Renormalised vacuum expectation values}

Renormalised v.e.v.s can be calculated by replacing $\alpha_F$ and $\beta_F$ in \eqref{eq:vevs}
by the differences $\alpha_{\text{reg}}^{\text{Had}} = \alpha_F - \alpha^{\text {Had}}_{\text {div}}$ and
$\beta_{\text{reg}}^{\text{Had}} = \beta_F - \beta^{\text{Had}}_{\text {div}}$:
\begin{subequations}
\begin{align}
 \braket{\psibar\psi}^{\text{Had}}_{\text{ren}} =& -\frac{\omega^3}{4\pi^2}\left(1 - \frac{k}{6} -k^2 + k^3\right)\nonumber\\
 &+ \frac{\omega^3 k (k^2 - 1)}{2\pi^2} \left[\psi(k) +
 \ln\left(\frac{e^{\gamma} \omega }{\nu_{\text{Had}} \sqrt{2}} \right)\right],\\
 \braket{J^\mu}^{\text{Had}}_{\text{ren}} =& 0,\\
 \braket{T^{\text{can}}}^{\text{Had}}_{\text{ren}} =& -\frac{\omega^4}{4\pi^2}\left(\frac{11}{20} + k - \frac{19k^2}{6} - k^3
 + \frac{5k^4}{2}\right)\nonumber\\
 &+ \frac{\omega^4 k^2(k^2 - 1)}{2\pi^2}
 \left[\psi(k) + \ln \left( \frac{e^{\gamma} \omega}{\nu_{\text{Had}} \sqrt{2}} \right)\right].
 \label{eq:had_tren_can}
\end{align}
\end{subequations}
When $k=0$, it can be seen that the trace anomaly obtained from \eqref{eq:had_tren_can} does not agree with \eqref{eq:traceanomaly}.
Furthermore, it is shown in \cite{art:dappiaggi09} that
the canonical definition \eqref{eq:Tmunu_sf} must be modified because the regularised propagator
$i S_{\text{reg}}^{\text{Had}}(x,x') = (\alpha_{\text{reg}}^{\text{Had}} + \beta_{\text{reg}}^{\text{Had}} \slashed{n})
\Lambda(x,x')$ does not satisfy
the Dirac equation. Hence, the renormalised v.e.v.~of
the canonical SET is, in general, not conserved. This is in contrast to Schwinger--de Witt renormalisation,  where both the
divergent and the finite parts of the SET are conserved, by construction \cite{art:christensen78}.
In~\cite{art:dappiaggi09} the conservation of the SET is restored by
changing the canonical definition of the SET,  adding a
term proportional to the Dirac Lagrangian multiplied by $g_{\mu\nu}$, as follows:
\begin{equation}
 T^{\text{new}}_{\mu\nu} = T_{\mu\nu}^{\text{can}} - \frac{1}{6} g_{\mu\nu}
 \left[\frac{i}{2} \overline{\psi} \slashed{D} \psi - \frac{i}{2} \overline{\slashed{D} \psi} \psi -
 m \overline{\psi} \psi\right].\label{eq:Tmunu_new}
\end{equation}
Since the Dirac Lagrangian vanishes when solutions of the Dirac equation are considered, $T^{\text{new}}_{\mu\nu}$
reduces to $T_{\mu\nu}^{\text{can}}$ in the classical (unrenormalised) case.
Taking the trace allows the new SET to be written in terms of the old one as:
\begin{equation}
 \braket{T^{\text{new}}}^{\text{Had}}_{\text{ren}} = \frac{1}{3} \braket{T^{\text{can}}}^{\text{Had}}_{\text{ren}} + \frac{2k \omega}{3}
 \braket{\psibar\psi}^{\text{Had}}_{\text{ren}}.
\end{equation}
The result is:
\begin{multline}
\label{eq:had_tren_new}
 \braket{T^{\text{new}}}^{\text{Had}}_{\text{ren}} = -\frac{\omega^4}{4\pi^2}\left(
 \frac{11}{60} + k - \frac{7k^2}{6} - k^3 + \frac{3k^4}{2}\right)\\
 + \frac{\omega^4 k^2(k^2 - 1)}{2\pi^2} \left[\psi(k) + \ln \left( \frac{e^\gamma \omega}{\nu_{\text{Had}} \sqrt{2}}\right)
  \right].
\end{multline}
Setting $k=0$ in \eqref{eq:had_tren_new} gives the expected trace anomaly \eqref{eq:traceanomaly}.
The above result can be compared to that obtained using zeta-function regularisation \cite{art:camporesi92}:
\begin{multline}
 \braket{T}_{\text{ren}}^{\zeta} = -\frac{\omega^4}{4\pi^2}\left(
 \frac{11}{60} + k - \frac{7k^2}{6} - k^3 + \frac{3k^4}{2}\right)\\
 + \frac{\omega^4 k^2(k^2 - 1)}{2\pi^2} \left[\psi(k) + \ln \left( \frac{\omega}{\nu_{\zeta}}\right)\right],
 \label{eq:zeta_tren}
\end{multline}
where $\nu_\zeta$ is an arbitrary renormalisation mass scale.
Our result \eqref{eq:had_tren_new} obtained using Hadamard renormalisation is in excellent agreement with the
zeta-function regularisation result
above, provided that the two  renormalisation mass scales are related by:
\begin{equation}
 \nu_\zeta = e^{-\gamma} \nu_{\text{Had}} \sqrt{2}.
\end{equation}
This relationship between the renormalisation mass scales for zeta-function and Hadamard renormalisation matches
that found in the case of a quantum scalar field on $n$-dimensional adS \cite{art:kent14,art:moretti99}.

\section{Discussion and conclusions}

In this letter, we have studied the renormalised vacuum expectation values (v.e.v.s) of the fermion condensate (FC),
charge current (CC) and stress-energy tensor (SET) for a quantum
fermion field of mass $m$ on four-dimensional anti-de Sitter (adS) space-time.
Due to the maximal symmetry of the space-time, we have been able to use a geometric approach to derive closed-form expressions
for the Feynman propagator and the renormalised v.e.v.s.
We used two methods of regularisation, namely the Schwinger--de Witt approach \cite{art:christensen78} and Hadamard
renormalisation \cite{art:najmi84,art:dappiaggi09}, comparing these, respectively, with previously published results
using Pauli-Villars and zeta-function renormalisation \cite{art:camporesi92}.
The v.e.v.~of the SET computed using Schwinger--de Witt renormalisation agrees with that found using Pauli-Villars;
and the v.e.v.~of the SET from Hadamard renormalisation agrees with that from the zeta-function approach, in each
case providing there is a relationship between the relevant renormalisation mass scales.

Wald's axioms \cite{art:wald77} uniquely define the v.e.v. of the SET up to a local conserved tensor.
Since the v.e.v. of the SET computed here is a constant multiplied by the metric tensor, whatever the renormalisation prescription,
the difference in v.e.v.s of the SET computed using Schwinger-de Witt and Hadamard renormalisation is trivially a local conserved tensor.
The agreement between Pauli-Villars and Schwinger-de Witt renormalisation is not surprising; both methods isolate the purely geometric divergent terms in the
Feynman propagator by using a large-mass expansion.  Their equivalence was shown for a scalar field on two-dimensional space-time in \cite{art:bunch78}.
For a quantum scalar field, the equivalence of the zeta-function and Hadamard renormalisation methods is proven in \cite{art:moretti99} and we would expect a similar result to hold for
a quantum fermion field.
For a quantum scalar field, the Schwinger-de Witt representation of the Feynman propagator is a special case of the Hadamard representation with additional, finite, renormalisation terms
\cite{art:dec06,art:hack12}.
Here, we have used an auxiliary propagator which satisfies a Klein-Gordon-like equation.
Therefore, the additional finite renormalisation terms in the Schwinger-de Witt method compared with the Hadamard method lead to the discrepancy in the corresponding final v.e.v.s.

We now compare our results using the two approaches to renormalisation.
In both approaches the v.e.v.~of the CC vanishes identically. The v.e.v.~of the FC differs in the two
approaches if the mass of the fermion field $m=k\omega $ is nonzero:
\begin{equation}
 \braket{\psibar\psi}_{\text {ren}}^{\text{Had}} -  \braket{\psibar\psi}_{\text{ren}}^{\text{SdW}}  =
 -\frac {\omega ^{3}k^{3}}{4\pi ^{2}} +\frac{\omega^3 k (k^{2} -1)}{2\pi^2}
 \ln \left( \frac {e^{\gamma} \nu _{\text {SdW}}}{\nu _{\text {Had}}\sqrt{2}} \right) .
\end{equation}
For the v.e.v.~of the SET, when $k=0$, using either Schwinger--de Witt or Hadamard renormalisation we find the expected trace
anomaly for massless fermion fields \eqref{eq:traceanomaly}.
We note that the trace anomaly is negative, as was the case for a quantum scalar field on adS \cite{art:kent14}.
When $k\neq 0$, the two approaches to renormalisation do not yield the same
answer for the trace:
\begin{multline}
 \braket{T^{\text{new}}}_{\text{ren}}^{\text{Had}} -  \braket{T^{\text{can}}}_{\text{ren}}^{\text{SdW}}\\
 = \frac {\omega ^{4}}{4\pi ^{2}}
 \left(k^{2}-\frac {3k^{4}}{2}\right)
 +\frac{\omega^4 k^{2} (k^{2} -1)}{2\pi^2}
 \ln \left( \frac {e^{\gamma} \nu _{\text {SdW}}}{\nu _{\text {Had}}\sqrt{2}} \right).
\end{multline}
For small values of the fermion mass, the trace of the v.e.v.~of the SET is negative for
both Schwinger--de Witt and Hadamard renormalisation.
As $m$ increases, the exact behaviour of the trace $\braket {T}$ depends on the value of the renormalisation
mass scale and the renormalisation scheme chosen.
There will typically be at least one value of $m$ for which the trace  vanishes, similar to the behaviour seen
for a quantum scalar field \cite{art:kent14}.
For sufficiently large $m$, the trace is always positive for both renormalisation schemes as the term
$\sim k^{4}\psi (k)$
in \eqref{eq:sdw_tren} and \eqref{eq:had_tren_new} becomes dominant.
We note that, as seen for a quantum scalar field \cite{art:kent14}, the v.e.v.~of the SET grows without bound as
the mass of the fermion field gets very large.
This counter-intuitive result is due to the negative curvature of adS space-time.

Due to the maximal symmetry of both the underlying space-time and the global adS vacuum, the v.e.v.~of the SET is
proportional to the metric tensor
$g_{\mu \nu }$ regardless of the renormalisation method employed.  If we consider the back-reaction of the quantum
fermion field on the geometry, this is governed by the semi-classical Einstein equations. As with a quantum scalar
field in the global adS vacuum \cite{art:kent14}, the semi-classical Einstein equations in this case are readily
solved simply by making a one-loop quantum correction to the cosmological constant.

As can be seen most easily in the Hadamard approach to renormalisation, the short-distance singularities of the
fermion Feynman propagator are independent of the choice of quantum state. Therefore, since we have now computed
renormalised expectation values when the quantum fermion field is in the global adS vacuum state, renormalised
expectation values for other quantum states of the fermion field can readily be computed by finding differences
in expectation values between two quantum states.  We will apply this method in a forthcoming publication
\cite{art:veaew}, where we consider thermal states for a massive fermion field on adS.

\section*{Acknowledgments}
V.E.A.~was supported by a studentship from the School of Mathematics and Statistics at the University of Sheffield.
The work of E.W.~is supported by the Lancaster-Manchester-Sheffield Consortium for
Fundamental Physics under STFC grant ST/L000520/1.

%\section*{References}

\end{document}